\documentclass[preprint2]{aastex}

\input psfig.sty

\newcommand{\chg}[1]{#1}
 
\slugcomment{Astronomical Journal in press}
\shorttitle{{\it ASCA/ROSAT} Observations of Holmberg II.}
\shortauthors{Miyaji, Lehmann \&  Hasinger}
\begin{document}
\title{Multiple Components of the Luminous Compact X-ray Source 
   at the Edge of Holmberg II observed by {\it ASCA} and {\it ROSAT}}

\author{Takamitsu Miyaji}
\affil{Department of Physics, Carnegie Mellon University \\
    Pittsburgh, PA 15213 (miyaji@astro.phys.cmu.edu)}

\author{Ingo Lehmann\altaffilmark{1}}
\affil{Department of Astronomy and Astrophysics, 
   Pennsylvania State University \\ 
   University Park, PA 16802 (ilehmann@aip.de)}

\and

\author{G\"unther Hasinger}
\affil{Astrophysikalisches Institut Potsdam, An der Sternwarte 16\\
  D-14482, Potsdam, Germany  (ghasinger@aip.de)}

\altaffiltext{1}{Also Astrophysikalisches Institut Potsdam (present address)}

\begin{abstract}
 We report the results of the analysis of new {\it ASCA} observations and 
archival {\it ROSAT} data of the compact luminous X-ray source found at 
the edge of the nearby star-forming dwarf galaxy  \objectname[]{Holmberg II}
(\objectname[]{UGC 4305}) in the \objectname[]{M81} group. We have found 
a number of new features in the X-ray properties of this source. Our new 
{\it ASCA} spectrum revealed that the X-ray emission extends to the hard band 
and can be best described by a power-law with a photon spectral index 
$\Gamma \sim 1.9$ \chg{while a $kT\sim 5$ [keV] thermal plasma with a low 
abundance ($\sim 0.2 Z_\sun$) is also acceptable.}  The {\it ASCA} spectrum 
does not fit with  a multi-color disk blackbody, unlike some off-nucleus 
X-ray sources with similar luminosities. The joint {\it ASCA}-{\it ROSAT} 
spectrum suggests two components to the spectrum: the hard power-law 
component and a warm thermal plasma ($kT\sim 0.3$ [keV]). An additional 
absorption over that of our galaxy is required.
 The wobble correction of the  {\it ROSAT} HRI image  
has clearly unveiled  the existence of an extended component 
which amounts to $27\pm 5\%$ of the total X-ray emission.


 These observations indicate that there are more than one component
in the X-ray emission. The properties of the point-like component
is indicative of an accretion onto an intermediate mass blackhole,
\chg{unless a beaming is taking place}. 
We argue that the extended component does not 
come from electron scattering and/or reflection by scattered 
optically-thick clouds of the central radiation. 
Possible explanations of this X-ray source include multiple 
supernova remnants feeding an intermediate-mass blackhole. 
   
\end{abstract} 
 



\section{Introduction}
 
 Off-nuclear (non-AGN) point-like luminous X-ray sources, 
which exceed the Eddington luminosity of a usual 
stellar-mass compact object like neutron stars and galactic binary 
blackholes ($L_{\rm x}\gtrsim 10^{39}$ $[{\rm erg\,s^{-1}}]$) 
in nearby galaxies have been recognized in many galaxies
\citep{fabbiano,col_mus,read}. The luminosity distribution of 
these sources extends up to a few $\times 10^{40}$ $[{\rm erg\,s^{-1}}]$. 
The most luminous one has been found in the nuclear region of the starburst
galaxy \objectname[]{M82} \citep{ptak,matsumoto1} reaching
$\sim 10^{41}$ $[{\rm erg\,s^{-1}}]$. The location of this
source was recently verified to be offset from the dynamical 
center of the galaxy by {\it Chandra} \citep{matsumoto2}.
  
 One interpretation of these X-ray sources is an accretion onto 
intermediate-mass blackholes of $10^{2-3}$ $M_\sun$, while there are 
other possibilities including beaming, super-Eddington accretion, and 
a young supernova remnant in a dense interstellar medium.  Probably 
they do not form a single class of X-ray sources.  
The spectra of many of these sources can be described by a multicolor 
disk blackbody model with or without additional
hard tail \citep{makishima,col_mus}. The inner disk temperatures 
of the sources studied by \citet{makishima} are  $kT\sim 1.1-1.8$ [keV]. 
Since these temperatures are higher than that from an
intermediate ($10^{2} M_\sun$) blackhole accretion, they proposed
a picture of a spinning (Kerr) blackhole with a much smaller 
inner-disk radius than that of a non-spinning (Schwarzschild) one.

 The luminous X-ray source in the star-forming dwarf galaxy Holmberg II 
(UGC 4305) is particularly peculiar among  these cases in terms of the nature
of the host galaxy and the location of the X-ray source. The galaxy 
is nearby at a distance of 3.05 Mpc, which was derived
from Cepheid observations \citep{hoessel}. It is a  dwarf irregular 
galaxy in the M81 Group with an on-going star-formation activity.  Like 
in other galaxies of this class, numerous HII regions are scattered 
throughout the galaxy \citep{hodge}.  The bright
X-ray source, which has been cataloged in the {\it ROSAT} Bright
Survey (RBS) \citep{rbs}, is one of the most luminous X-ray source 
seen in dwarf galaxies ($L_{\rm x}\sim 10^{40}$ ${\rm [erg\,s^{-1}]}$).

 It was observed three times with {\it ROSAT} PSPC as
pointed observations and once with HRI. \citet{zezas} (hereafter
ZGW) studied the {\it ROSAT} data on this X-ray source. Their main
conclusions were: 1) The X-ray source seems point-like and located near the 
edge of the galaxy. Its position is consistent with one of the compact HII 
region complexes of the galaxy considering an uncertainty of $\sim 6$ 
arcseconds for the {\it ROSAT} absolute astrometry from its star tracker.
A chance spatial co-incidence of this HII-region to an unrelated X-ray 
source of this brightness is highly unlikely ($\sim 3\; 10^{-6}$). 
2) The source is time variable on scales of days and years. 
3) The spectrum is soft and can be fitted with a steep power-law 
($\Gamma \sim 2.7$) or a thermal plasma 
(Raymond-Smith) of $kT\sim 0.8$ keV. Time variability
suggests an accretion onto a compact object, but the luminosity
is 2-4 orders of magnitude larger than typical X-ray binaries.

 There is a radio-knot which is co-spatial with the HII-region
in the X-ray source region. It is one of the three showing radio 
spectra suggestive of a non-thermal emission \citep{tongue}, 
suggesting a recent supernova or multiple supernovae. However, 
currently there is no indication of a special character in 
the optical-radio data in this particular HII-region which is 
presumably associated with the compact X-ray source compared 
with other knots in the galaxy. 
 
  As a part of our study of a sample of apparent non-AGN
galaxies which have compact luminous X-ray sources, selected from the
RBS, we have made further X-ray studies of Holmberg II. In particular, 
we have obtained a 20 [ks] {\it ASCA} observation of this X-ray
source in order to search for the hard X-ray signature of an AGN-like 
activity. In this paper, we report the results from our data analysis
of our {\it ASCA} observation and of the {\it ROSAT} data
from archives. 

The scope of the paper is as follows. In Sect. \ref{sec:obs}, we summarize
the {\it ASCA} and {\it ROSAT} observations and data used
in our analysis.  In Sect. \ref{sec:spec} we report our
new spectral analysis of the X-ray source by a joint 
{\it ASCA}-{\it ROSAT} PSPC analysis. In Sect. \ref{sec:hri}, 
we searched for an extension by applying a wobble-correction to the 
{\it ROSAT} HRI data. Long and short-term X-ray light curves are 
examined in Sect. \ref{sec:lc}. Finally the results are discussed 
in Sect. \ref{sec:disc}.

\section{Observations}\label{sec:obs}
 
 Holmberg II was observed with {\it ROSAT} five times. 
Firstly, it was bright in the {\it ROSAT} All-Sky Survey
(RASS) \citep{rassb} and cataloged in the RBS \citet{rbs}. It was
observed three times with {\it ROSAT} PSPC pointed observations
in 1992-1993 for 3.7-11 [ks] each. Also it was observed with 
{\it ROSAT} HRI for 7.8 [ks]. The {\it ROSAT} pointed data have been 
retrieved from the High Energy Astrophysics Archival Research Center
(HEASARC) located at NASA Goddard Space Flight Center. 

 The ASCA observation of Holmberg II has been made with a pointing at 
the X-ray source in 1999 as one of the accepted targets from
our proposal in the {\it ASCA} EAO-6 (AO-7) program (PI=Lehmann). 
The brightness of the source is such that we can use the SIS FAINT 
mode for both the medium and high bit rates with the 1-CCD mode. 
Thus we chose to use the 1-CCD/FAINT modes throughout. 
For the subsequent analysis, we use the BRIGHT2 mode data converted 
from the FAINT mode. For GIS, we have used the  PH mode as usual.  
The log of {\it ROSAT} and {\it ASCA}
pointed observations is shown  in Table \ref{tab:log}.

\placetable{tab:log}
 
\section{X-ray Spectral Analysis using {\it ASCA - ROSAT} PSPC}
\label{sec:spec}
\subsection{Extracting  {\it ASCA} and {\it ROSAT} spectra}

 Extraction and preparation of spectra and response matrices have 
been made using tools included in FTOOLS 5.0 and subsequent 
spectral fittings have been made using XSPEC 11.
 
 The source spectra have been extracted from event files which have been
screened using the standard screening criteria. The source-spectrum
extraction radii were 6 and 3 arcminutes for the GIS and 
SIS respectively. The background spectra have been extracted from
off-source areas from the same detector, excluding 7.5 and  
and 3.7 arcminutes from the X-ray source for the GIS and SIS data 
respectively. The background subtraction have been made 
by scaling the background spectra by the areas of extraction, 
as automatically made by
XSPEC. Because of the brightness of the source, the analysis is little 
affected by the uncertainties related to background scalings. The two 
GIS and two SIS spectra are co-added respectively. The background 
spectra and response files have been generated respectively for the 
summed GIS 2+3  and SIS 0+1 spectra. 

 The {\it ROSAT} PSPC spectra have been extracted using an extraction
radius of 1 arcminute. Background spectra have been extracted from
an annular region around the source, between 1.25 and 3.75 arcminutes.
One spectrum has been made for each of the three PSPC pointed 
observation (Table \ref{tab:log}). We see no evidence of a spectral
change among these three datasets (see below). Thus the three 
spectra for these three sequences have been co-added for the
joint analysis with the {\it ASCA} data. 
 
 The {\it ASCA} and {\it ROSAT} spectra have been re-binned such that 
each bin contains at least 25 counts in order to utilize the 
$\chi^2$ statistics during the spectral analysis. The source was 
bright enough for this re-binning process without sacrificing the 
resolution.    

\subsection{Spectral Analysis}
\label{sec:specana}

 The channel energy ranges used for the analysis are 0.15-2 keV, 
0.6-10 keV, and  1.0-7 keV for the {\it ROSAT} PSPC, {\it ASCA} GIS,
and {\it ASCA} SIS, respectively. Because of the calibration uncertainties
related to the radiation damage, we have ignored 
{\it ASCA} SIS channels below 1 keV. 
 
 Firstly,  we have verified the basic fitting results
of ZGW with joint fits to the three PSPC spectra,
where all model parameters are joined except for the global
normalization. As observed by ZGW, slight variations
of the global normalization have been observed. This is discussed
in Sect. \ref{sec:lc}. When all parameters are separately
fit for the three datasets, the best fit parameters are always
consistent with the joint ones within errors. This warrants 
that we can use the summed spectra for {\it PSPC} for further 
analysis.

\begin{figure}
\plotone{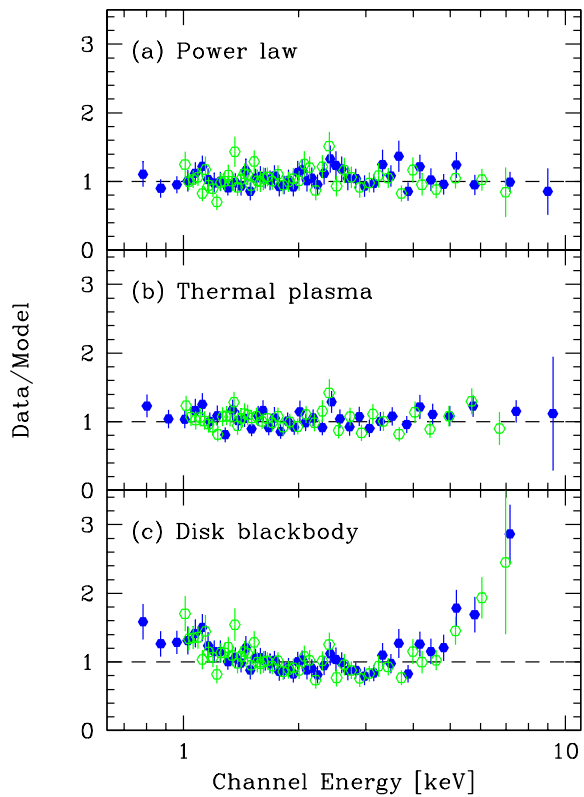}
\caption
 {The residuals for three model fits of the {\it ASCA} GIS (filled
 circles)/SIS (open circles) 
 spectra are shown in terms of the ratio of the data and model.
 The models are (a) a single power-law, (b) a thermal plasma, 
 (c) a disk blackbody. See Table \ref{tab:spec} 
 for the best-fit parameters. 
 \label{fig:ascaresid}}
\end{figure}

\placetable{tab:spec}
\placefigure{fig:ascaresid}

 Next we have made a spectral analysis of {\it ASCA} only (GIS \& SIS).
We have tried to fit the spectra with three models: (A1) 
a single power-law, (A2) a thermal
plasma using the XSPEC {\it mekal} model, and (A3) a disk 
blackbody \citep{mitsuda} (XSPEC model {\it diskbb}), where ``A''
stands for {\it ASCA}.  The best-fit parameters are shown in 
Table \ref{tab:spec} and fit residuals are shown in 
Fig. \ref{fig:ascaresid}.  The single power-law of photon index 
$\Gamma \sim 1.9$ gave the best fit without inclusion of any absorption. 
 The photon indices for separate fits to the
gave $\Gamma = 1.85\pm 0.07$ and $1.87\pm 0.07$ for GIS and SIS,
respectively ($N_{\rm H}$ fixed to 0). The agreement verifies 
the goodness of our background subtraction.
The 90\% upper-limit to the column density of the absorbing gas was 
$N_{\rm H}\le 1.3\;10^{21}$  $[cm^{-2}]$. 
The thermal plasma model with $kT\sim 5$ [keV] also gave  a good fit 
with $\chi^2/\nu=0.84$. \chg{We have obtained a heavy element abundance of
$Z=0.2\pm 0.2 Z_\sun$, which is marginally consistent with the O/H 
ratio of $\sim 0.4$ (relative to solar) for this galaxy \citep{hun_gal85}.
However note that the metallicity of the source of X-ray emission 
does not necessarily be equal to that of some average for the HII 
regions scattered throughout the galaxy and that the hard X-ray mainly
measures the iron abundance, which does not necessarily match with that
of oxygen.}

The multicolor disk blackbody model failed to fit the {\it ASCA}
data as shown in Fig. \ref{fig:ascaresid}(c). Thus the disk-blackbody
component does not dominate the hard X-ray emission observed by
our {\it ASCA} observation, unlike the cases
of, e.g., \objectname[]{IC 342} source 1 and \objectname[]{NGC 1313} 
source B studied by \citet{makishima}. The power-law index
of the {\it ASCA} spectrum is much harder than that of  {\it ROSAT}
PSPC spectrum ($\Gamma\approx 2.7$; ZGW), which has response in
$E\la 2$ keV. 

\begin{figure}
\psfig{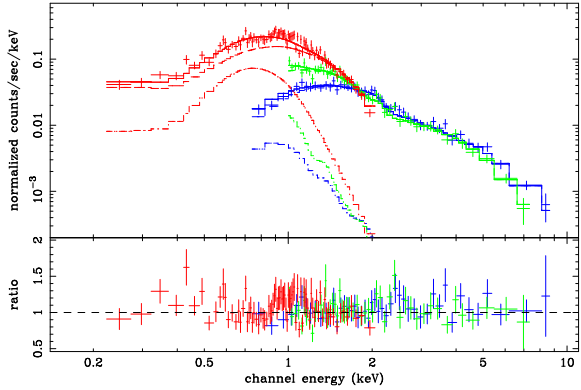}
\caption
  {The {\it ASCA} GIS (0.6-10 keV)/SIS (1-7 keV) and 
  {\it ROSAT} PSPC (0.2-2 keV) spectra are shown with folded model 
  predictions for the best-fit 
  power-law plus thermal (MEKAL) model with absorption (Fit J1).  
  The fit residuals are shown in the lower panel in terms of the ratio
  of the data and the model. The error bars show 1$\sigma$ errors.
  \chg{The detector correponding to  each spectrum can be identified by 
  the energy range. The power-law (dashed) and thermal (dot-dashed) 
  components of the folded model have also been shown separately for 
  each instrument.}     \label{fig:spec}}
\end{figure}

\placefigure{fig:spec}

  Finally, we have made a joint spectral analysis using  all the 
{\it ROSAT} PSPC and  {\it ASCA} GIS/SIS data. We have summed all 
the three PSPC spectra for the joint spectral fits. We have joined 
all the model parameters except for 
the global normalization as before.  Because the {\it ROSAT} spectrum, 
covering a lower energy range, is much softer than that of {\it ASCA}, we 
have tried models with hard power-law ($\Gamma\sim 1.9$) with 
some soft excess and an absorbing column. Adding a softer power-law 
component did not give a satisfactory fit. As the soft component, we 
have tried the thermal plasma and disk blackbody models. As shown in 
Table \ref{tab:spec}, both ``Power-law+Thermal'' (J1) and 
``Power-law+Disk blackbody'' (J2)
models gave reasonable fits. \chg{Note that the abundance of
the thermal plasma have been fixed to 0.4 (see above) for J1. 
When this was a free parameter, the fit could not constrain
it.} In both J1 and J2, the absorbing column
was $N_{\rm H}\sim 0.7-1\;10^{21}$ ${\rm [cm^{-2}]}$, which 
exceeds the galactic value implied by the 21cm data  of
$N_{\rm H}\sim 3\;10^{20}$ \citep{21cm}, indicating an absorbing
component within Holmberg II. The soft excess component,
which can either be described as a $kT\sim 0.3$ [keV] thermal plasma 
or a  $kT_{\rm in}\sim 0.2$ [keV] disk blackbody spectrum, has
20-30\% of the total 0.5-2 keV total luminosity. 
The {\it ROSAT} and {\it ASCA} spectra are shown with folded models 
and residuals in Fig. \ref{fig:spec} for J1. 
The total luminosity of the X-ray source is $\approx 1.0\; 10^{40}$ 
$[{\rm erg\,s^{-1}}]$ in the 0.1-10 keV after correcting for
absorption (GIS normalization). This corresponds to an Eddington
luminosity of $\sim 80 [M_\sun]$. 

 For the fit J2, the GIS normalization of the disk blackbody  
component gives $r_{\rm in}\sqrt{\cos \theta} = 3500-7500$ 
$[km]$, where $r_{\rm in}$ is the inner disk radius, $\theta$ is the
viewing angle and the 90\% confidence error was searched for by 
allowing all other variable parameters to vary. 
Assuming $r_{\rm in}=3\,R_{\rm S}$, where
$R_{\rm S}$ is the Schwarzschild radius, this corresponds
to a blackhole mass of 10-25 $(\cos \theta)^{-0.5} [M_\sun]$, while 
\citet{col_mus} pointed out that the $r_{\rm in}$ estimation based 
on the disk blackbody within XSPEC is likely to be underestimated.
Further discussion is made in Sect. \ref{sec:disc}
       
\chg{We have also tried a two-thermal plasma model fit (J3), where 
the relative abundances of two thermal plasma have been  fixed to equal.
This model gave an acceptable fit but with a hard thermal component
(Thermal$_{\rm h}$) with $kT_{\rm h}\sim 6$ [keV] and a soft thermal
component (Thermal$_{\rm s}$) with $kT_{\rm s}\sim 0.3$ [keV]. In this case,
the fitted abundance was very low $Z=0.02^{+.03}_{-.01}$.
We did not find a satisfactory fit for the model where 
the hard component is a thermal plasma and the soft component is 
a power-law.}   
 
 One caveat on our joint fitting analysis is a  possible
cross-calibration problem. \citet{iwas_rosasca} found that 
the {\it ROSAT} PSPC spectrum is much softer than that of 
{\it ASCA} SIS for the simultaneous observations of NGC 5548
($\Gamma=2.35$ for {\it ROSAT} PSPC versus $\Gamma=1.95$ for
{\it ASCA} SIS). This might have been partially be caused by 
the soft excess combined with the spectral resolution of PSPC
or there may be real cross-calibration problems. The origin of
this discrepancy is inconclusive. In our case, the difference 
of $\Delta \Gamma\sim 0.7$ between the {\it ROSAT} PSPC and
{\it ASCA} GIS spectra is much larger than that \citet{iwas_rosasca}
observed for a similar {\it ASCA} index. Thus there certainly is  
a soft component over the extrapolation of the hard power-law even 
in case the discrepancy observed by \citet{iwas_rosasca} was 
solely caused by calibration problems.

\section{{\it ROSAT} HRI Wobble-Correction and Spatial Extension} 
\label{sec:hri}

\begin{figure}
\plotone{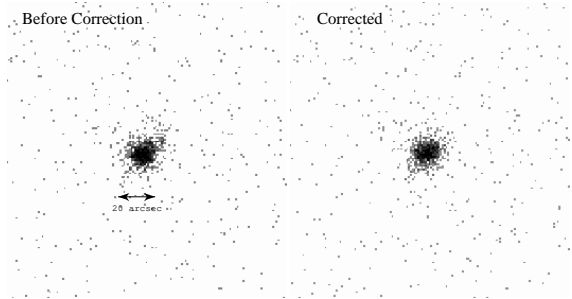}
\caption
{The {\it ROSAT} HRI image of Holmberg II before (north is up, east is left)
 and after the wobble aspect correction. Note that a faint knot seen
 at the north-west in the original image have disappeared after 
 the correction.    
 \label{fig:wcorr}}
\end{figure}

 There are wobble-phase dependent systematic errors with the aspect 
solution for the {\it ROSAT}, which could lead to a detection of 
spurious extended component in the HRI data \citep{harris98,morse}.
This has lead to ZGW make a conservative conclusion
that they had detected no extension in the HRI image. 
We have applied a correction for this effect by creating images 
in 10 wobble-phases, centering each of these images independently, 
and re-constructing the image using the centering information
\citet{lehmann}. The original and corrected HRI images
are shown in Fig. \ref{fig:wcorr}. The faint knot seen in the
NW in the original HRI image has disappeared in the corrected image.
This reconstruction  worked very well down to an HRI count rate 
of $\sim 0.1$ [cts s$^{-1}$] for a large number of stars, but
can be even used for fainter sources \citep{crawford}.

\placefigure{fig:wcorr}

\begin{figure}
\psfig{file=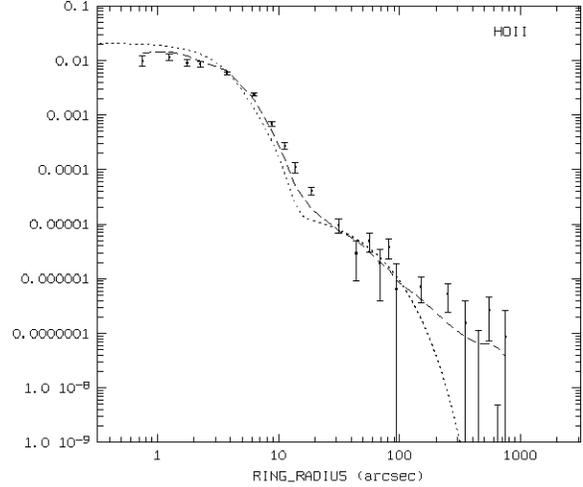,angle=270,width=\hsize,clip=}
\caption
{\chg{The wobble-corrected radial profile of the compact X-ray source in Holmberg II 
 is compared with the theoretical HRI PSF (dotted line) and the re-calibrated 
 HRI PSF (dashed line). The bars mark $1\sigma$ errors of the data points.
 An extended component is clearly seen in $\gtrsim 10$ arcseconds.}
 \label{fig:hriprof}}
\end{figure}

\chg{The radial profile of the wobble-corrected HRI image has been compared
with the theoretical HRI PSF from \citet{hri} and the 
re-calibrated HRI PSF in Fig. \ref{fig:hriprof}. The re-calibrated HRI PSF was 
derived from wobble-corrected HRI images of 21 stars from the RBS 
\citet{rbs}. The re-calibrate HRI PSF shows a slightly deviation 
from the theoretical PSF in the radial range between 10 and 30 arcseconds. 
The comparison of the radial profile with both PSF's indicates an 
extended component. Subtracting 
the re-calibrated PSF, $27\pm 5$\% (1$\sigma$ error) of the 
X-ray emission is in the extended component and most of the excess comes from 
a radius of $\sim 10\arcsec$, corresponding to $\sim 150$ [pc].}

\placefigure{fig:hriprof}
  
\section{Variability}
\label{sec:lc}

\begin{figure}
\plotone{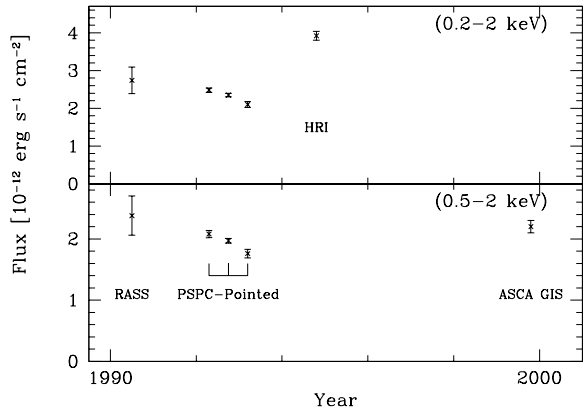}
\caption
{The long-term light curves of 
 Holmberg II from 1990-1999 in the 0.2-2 keV ({\it ROSAT}
 PSPC and HRI) and 0.5-2 keV ({\it ROSAT} PSPC and {\it ASCA}
 GIS) bands. The error bars are 1$\sigma$. \label{fig:llcurve}}
\end{figure}

 The {\it ASCA} GIS flux in the 0.5-2 keV band calculated using the spectral 
model J1 have been compared with the previous {\it ROSAT} observations.
Because {\it  ASCA} GIS is sensitive only above 0.6 keV, while {\it ROSAT}
HRI has no spectral resolution and sensitive in 0.2-2 keV, we have 
made separate long-term light curves for the 0.2-2 and 0.5-2 keV bands. 
These are shown in Fig. \ref{fig:llcurve}. The fluxes from 
{\it ROSAT} are slightly different from Fig. 3. of ZGW probably
because of the difference in spectral models.  As seen in Fig. 
\ref{fig:llcurve}, there is a
30\% decrease in flux between the first (April 1992) and third 
(March 1993) PSPC pointed observations and a factor of $\sim 2$ increase 
towards the HRI observation (Oct, 1994). The source brightness came back 
to a flux close to the PSPC observation during the {\it ASCA} 
observation in 1999. This does not necessarily mean that the source 
is variable at a time scale of ``years'', rather it means that there
is a variability at a timescale longer than the elapsed
time of a single observation ($10^5-10^6$[s]). 
       
\placefigure{fig:llcurve}


\chg{We also searched for evidence of time variability in the {\it ASCA}
data. First, we have defined the combined Good Time Intervals  (GTI),
which consist of the intersection of GTI's from all the 
SIS and GIS detectors. We have used channels corresponding to 
0.6-10 keV for the GIS's and 0.7-7 keV
for the SIS's respectively. For each detector, binned source and
background light curves have been extracted using a 1024 second binsize.
The extraction regions are the same as those of  the spectral analysis. 
For each time bin, only the interval overlapped with the combined GTI 
have been used to calculate the countrate. The background-subtracted light 
curves have been co-added to make the final light curve. There is no 
significant correlation between the background-subtracted source light 
curve and the background light curve. This also warrants goodness of 
the background subtraction.}  

\begin{figure}
\plotone{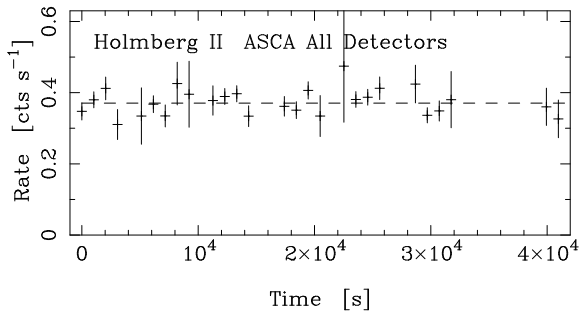}
\caption
{\chg{The {\it ASCA} light curve of 
 Holmberg II with 1$\sigma$ errors. Background-subtracted
 source counts for the all four detectors have been co-added.
 The energy ranges are 0.6-10 and 0.7-7 keV for the GIS and
 SIS respectively. 
 The dashed line shows the best-fit constant value.} 
 \label{fig:ascalc}}
\end{figure}

\chg{The resulting light curve is shown with the best $\chi^2$-fit constant
value in Fig. \ref{fig:ascalc}. No point is more than 2$\sigma$ away from 
the constant value and $\chi^2/\nu=22.2/26$. Thus  there is no 
evidence for variability at time scales between $10^3$ and $10^4$ 
seconds in the {\it ASCA} data. Separate light curves for the
hard ($E\ge 2$ [keV]) and soft ($E<2$ [keV]) bands did not show any sign
of variability either. This is in contrast to the {\it ROSAT}
light curves reported by ZGW, which show a convincing case of 
gradual decrease of flux by a factor of $\sim 3$  over $2\times 10^5$ 
seconds during the second pointed PSPC observation and less 
convincing indications of shorter timescale variabilities.}

\placefigure{fig:ascalc}

\section{Discussion}
\label{sec:disc}

 Our analysis suggests at least two-components of the X-ray
emission. An extended and a point-like component have been revealed 
by the spatial analysis. The variability also provides evidence for
at least a compact component. The joint {\it ASCA}/{\it ROSAT}
spectral analysis suggests a non-thermal power-law component plus 
a warm thermal component or a moderately low temperature disk blackbody
component. The overall spectrum can also be described as a superposition
of two thermal components with a very low metallicity. \chg{However,
the variability observed in this object argues against total
thermal origin of the X-ray source.}

  The X-ray source at the edge of Holmberg II shows unusual
characters, even given the large number of off-nuclear X-ray sources  
in nearby galaxies at similar luminosities. While most of the X-ray sources 
studied by \citet{col_mus} reside in the nuclear region, this is 
at the very edge of a dwarf galaxy.  Unlike a number of off-nuclear 
sources studied by \citet{makishima}, 
hard X-ray emission is not dominated by a disk blackbody component 
with $kT_{\rm in}\gtrsim 1$ [keV]. \chg{If the soft excess component 
comes from a disk blackbody emission, the implied blackhole mass 
from the normalization of this component is 10-25 
$(\cos \theta)^{-0.5} [M_\sun]$ assuming a Schwarzschild blackhole (see
Sect. \ref{sec:spec}). However, estimating the mass from the temperature
$kT_{\rm in}\sim 0.2$ [keV]  using Eq. (12) of 
\citet{makishima}, we obtain $\sim 10^4$ $[M_\sun]$. This large discrepancy
argues against the disk-blackbody interpretation of the soft excess,
even though we consider the underestimation of the mass pointed 
out by \citet{col_mus}, nearly edge on viewing angle, and spinning 
of the blackhole.}
 
 One of the most surprising results from our analysis is that these 
two (or more) bright X-ray components co-exist in a compact region 
in one of the numerous HII regions and no other X-ray source with 
a comparable brightness exists in other part of the galaxy. All other
X-ray sources scattered throughout the galaxy, which are  
either supernova remnants or X-ray binaries,
have luminosities of $L_{\rm x}\la 10^{37}$ ${\rm [erg\,s^{-1}\,cm^{-2}]}$ 
\citep{kerp}. Thus it is unlikely that the two (or more) components have 
independent origins. The possible  explanations of this X-ray source 
with observed multiple components are:
\begin{description}
\item[(a)] A supernova remnant (SNR) or a 
composite of SNRs (the $kT\sim$ 0.3 [keV] extended thermal 
component) feeds an intermediate mass blackhole with gas. 
\item[(b)]  \chg{An accretion onto a stellar-mass compact object,
e.g., similar to galactic blackhole X-ray binaries,
beaming towards our line of sight, is embedded in an SNR or a 
composite of  SNRs.}
\item[(c)] The non-thermal power-law and the soft excess components (possibly  
blackbody emission from the accretion disk around an intermediate-mass 
blackhole) come from the same compact region while the extended component is 
caused by scattering/reflection of the central source.  
\end{description}


 The flat radio spectrum \citep{tongue} in the region of
this X-ray source shows an existence 
of a supernova remnant or a composite of supernova remnants. 
This supports the possibilities
(a) or (b).  However, the extended component is too luminous for thermal 
emission from a single (or a few) supernova remnant(s) (SNR), 
considering its size of $\sim 200$ pc and its  0.5-2 keV 
luminosity of a few $\times 10^{39}$ $[{\rm erg\,s^{-1}}]$.
A single SNR can only be at this luminosity when it is 
very young ($\la 10^2$ years) \citep{schlegel94,schlegel99} and 
thus it cannot have grown into this scale size nor can feed the 
intermediate-mass blackhole located 200 [pc] away. Thus more
than a few supernova remnants are needed to make the extended
structure, which also feeds the blackhole causing the variable,
non-thermal emission. \chg{From our current data,  we are not
able  to distinguish whether the central source represents 
an accretion onto an intermediate-mass blackhole or a stellar mass
compact object (e.g. X-ray binaries) with beaming towards our line 
of sight.} 
 
 The possibility (c) is not likely for the following reasons.
The possibility that electron-scattering by surrounding diffuse gas 
is the origin of the extended emission has the following 
difficulty. Scattering gas with a column density of 
$N_{\rm H}\gtrsim 10^{23}-10^{24}$ $[{\rm cm}^{-2}]$ is required 
to account for the $\sim 30\%$ extended component. The gas would cause heavy 
photoelectric absorption unless the scattering diffuse gas is highly 
ionized \citep{wilson1068,elvis90}. If the gas was thermally ionized, 
the temperature of the gas must be greater than a few million 
degrees [K] in order that the absorption features due
to a column density of $N_{\rm H}=10^{23}$ $[{\rm cm}^{-2}]$ 
is not visible in the {\it ASCA} spectrum. 
For demonstration, we assume a uniform sphere of gas of 200 [pc] 
in radius filled with hot ionized gas with a column density
of $N_{\rm H}=10^{23}$ $[{\rm cm}^{-2}]$ from the central source. 
Such a sphere of thermal gas would produce X-ray emission 
which is several  orders of magnitude larger than observed.  
Photoionization also cannot be a valid explanation, because the 
lower limit of the ionization parameter required to be consistent
with the {\it ASCA} spectrum also requires several orders of magnitude
larger luminosity for the ionizing source.  We also consider 
a picture where the surfaces of optically-thick molecular clouds 
scattered throughout the region (giant molecular clouds where 
star forming activities are embedded) reflect the radiation 
from the central source, while our line of sight to the central 
source is not blocked by any of them. This picture also has a similar 
difficulty. In order to attain sufficient reflectivity in the 
{\it ROSAT} band ($E\la 2$) [keV], the surface of the clouds have 
to be highly ionized. Using the XSPEC model ``pexriv'' \citep{pexriv}, 
we estimate that an ionization parameter of $\xi\equiv L/(nR^2)>$ 
a few hundred $[{\rm erg\;cm\;s^{-1}}]$ is required to have the 
reflectivity of more than 20-30\% in this band.  
For a central ionizing source luminosity of $L \sim$ 
a few  $\times 10^{40}$ $[erg\,s^{-1}]$, at a distance of 
$R=200$ [pc], this ionization parameter corresponds to a gas 
density at the surfaces of the clouds of $n\la 10^{-3}$ $[{\rm cm}^{-3}]$. 
This is much lower than that of a usual interstellar medium 
and scattered clouds of this density cannot be optically thick.

\chg{If the intermediate-mass blackhole picture of this X-ray
source is the case, there is a question of its formation.}
\citet{taniguchi} discussed possible origins of the 
off-nucleus ``intermediate-mass'' blackholes
presumably responsible for off-nuclear luminous X-ray sources  
and their preferred model was multiple mergings of stellar 
remnants (stellar-mass blackholes and neutron stars).
\chg{\citet{matsushita} found a molecular superbubble in
the vicinity of a similar off-nuclear X-ray source in M82, suggesting
that an intensive star formation is connected to the formation of 
an intermediate-mass blackhole.} 
This may be the case  for the compact component of the X-ray source 
discussed here because of the intense star-forming activity in the region. 
However, the question remains as to why there is only one such X-ray source 
in this galaxy and not at the locations of other numerous HII regions. 
Probably the X-ray emission is a short-lived phenomenon, where an 
intermediate mass blackhole (similar things may exist in many star-forming 
regions) is fed by the passing of a shell of dense gas induced by multiple
supernova explosions.
 
 Chandra (scheduled as a GT target by Murray) and  XMM-Newton 
(GT target by Watson) observations of this source will 
certainly improve our knowledge of this peculiar X-ray source.
The Chandra observation can make spatially resolved 
X-ray spectroscopy and reveal whether the non-thermal component corresponds 
to the point-source and thermal to the extended component or 
it has a more complicated structure. 
The XMM-Newton observation would give a better spectral 
information with a much better statistics and much less cross-calibration 
uncertainties in the  $E\la 1$ keV region, where multiple components 
are apparent. It will also give variability information 
at short timescales.

\section{Conclusions}

 We summarize the main conclusions of our analysis:

\begin{enumerate}
\item New {\it ASCA} and archival {\it ROSAT} data of the 
 X-ray source in Holmberg II have been analyzed.  
\item The combined {\it ASCA} and {\it ROSAT} spectrum 
 shows can be described as a $\Gamma\sim 2$ power-law and
 a soft excess which can either be described as a $kT\sim 0.3$ [keV] 
 thermal plasma or $kT_{\rm in}\sim 0.2$ disk blackbody. We argue 
 against the disk blackbody interpretation of the soft excess 
 component based on the inconsistency between the luminosity and the 
 temperature. 
\item The wobble-corrected {\it ROSAT} HRI image indicates
 $\sim 25\%$ of the X-ray emission is extended at 
 a scale of $\gtrsim$10$\arcsec$ ($\sim $25\%). Variability
 in the {\it ROSAT} data shows that the point-like
 component probably comes from accretion onto a compact object.
\item It is remarkable that this multiple-component
 X-ray source is the only one with a similar strength
 among the numerous HII regions in the galaxy with a 
 similar nature. It is natural to suppose that 
 these two components are related. 
\item  Possible explanations of the nature of the 
 X-ray source include multiple supernova remnants (extended component) 
 feeding the accretion onto  an intermediate-mass blackhole. 
 \chg{We cannot, however, exclude the possibility that the central 
 source is a beamed radiation from an accretion onto a stellar mass
 (usual X-ray binaries) object.} We argue against a picture
 where electron scattering or reflection by scattered optically-thick 
 clouds of the central source makes the extended  X-ray emission. 
\end{enumerate} 

\acknowledgments

 This research have made use of data and software obtained from
the High Energy Astrophysics Archival Research Center
(HEASARC) located at NASA Goddard Space Flight Center. The 
authors appreciate the effort of {\it ASCA} and {\it ROSAT}
teams for having created and operated these superb 
observatories. The authors thank the referee, Hironori 
Matsumoto for his careful review and useful comments.










\begin{deluxetable}{cccc}
\footnotesize
\tablecaption{Log of {\it ROSAT} and {\it ASCA} Observations \label{tab:log}}
\tablewidth{0pt}
\tablehead{
\colhead{Detector(Mode)} & \colhead{Sequence}  & 
\colhead{Start Date} &
\colhead{Exposure [ks]} } 

\startdata
{\it ROSAT}  PSPC      & rp600140n00  & 14-APR-1992 &  7.3 \\
    ``        ``       & rp600431n00  & 29-OCT-1992 & 11.6 \\
    ``        ``       & rp600431a00  & 14-MAR-1993 &  3.7 \\ 
{\it ROSAT}  HRI       & rh600745n00  & 17-OCT-1994 &  7.8 \\
{\it ASCA}   GIS(PH)   & ad77075000   & 21-OCT-1999 &     
             16.5  \\
{\it ASCA}     SIS (1CCD/FAINT) & ``  &      ``       & 
       16.5(S0)/16.3(S1) \\
\enddata

\end{deluxetable}


\clearpage

\begin{deluxetable}{cl}
\footnotesize
\tablecaption{Results of the Spectral Analysis\label{tab:spec}}
\tablewidth{0pt}
\tablehead{
\colhead{Model} & \colhead{Parameters}\\
}
\startdata

\cutinhead{{\it ASCA} GIS \& SIS Data} 

{\bf PL} & $K_{\rm G}=1.0*$; $K_{\rm S}=.87\pm .05$; $\Gamma=1.87\pm 0.05$;\\ 
(A1)     & $F_{\rm p12}=2.5\pm 0.1$; ${\bf \chi^2/\nu=0.83}\,(170./206)$ \\ \\ 

{\bf Thermal} & $K_{\rm G}=1.0*$; $K_{\rm S}=.86\pm .05$; $kT = 4.8\pm 0.4$;\\ 
(A2)    & $F_{\rm t12}=2.2\pm .1$; $Z=0.2\pm 0.2$; 
	${\bf \chi^2/\nu=0.84}\,(208./206)$ \\ \\
        
Disk BB & $K_{\rm G}=1.0*$; $K_{\rm S}=.87\pm .05$; 
         $kT_{\rm in}=1.19\pm 0.05$; \\     
(A3) &   $F_{\rm d12}=1.9\pm .1$; ${\bf \chi^2/\nu=1.34}\,(276./206)$; \\ 

\cutinhead{ {\it ROSAT} PSPC \& {\it ASCA} GIS/SIS}\\

{\bf (PL+Thermal) $\cdot$ Abs.} & $K_{\rm G}=1.0*$; $K_{\rm S}=.87\pm .05$; 
             $K_{\rm P}=.79\pm .06$; $\Gamma=1.91\pm .04$;  \\
(J1)       & $F_{\rm p12}=2.6\pm .2$; $kT=.30\pm .05$;$Z=0.4*$; 
              $F_{\rm t12}=.67\pm .17$;\\
           &  $N_{\rm H20}=7.9\pm .06$; ${\bf \chi^2/\nu=1.02}\,(324./317)$\\ 
                         \\
{\bf (PL+Disk BB)$\cdot$ Abs.} & $K_{\rm G}=1.0*$; $K_{\rm S}=.86\pm .04$; 
             $K_{\rm P}=.78\pm .04$; $\Gamma=1.88\pm .07$;\\
(J2)       & $F_{\rm p12}=2.6\pm .2$; $kT_{\rm in} =.17\pm .02$; 
              $F_{\rm d12}=1.0\pm .2$\\
           & $N_{\rm H20}=9.6\pm 1.1$; ${\bf \chi^2/\nu=1.04}\,(329./317)$ \\        
                      \\
{\bf (Thermal$_{\rm h}$+Thermal$_{\rm s}$)$\cdot$ Abs.} & 
             $K_{\rm G}=1.0*$; $K_{\rm S}=.87\pm .05$; $K_{\rm P}=.79\pm .04$; 
             $kT_{\rm h}=5.9^{+1.4}_{-.9}$; \\
(J3)       & $F_{\rm th12}=2.0\pm .1$; $kT_{\rm s}=.35\pm .05$; 
             $F_{\rm ts12}=1.4\pm .2$;\\
           &   $Z=.02^{+.03}_{-.01}$; $N_{\rm H20}=8.5\pm 1.0$; 
              ${\bf \chi^2/\nu=1.00}\,(317./317)$ \\ 
\enddata


\tablecomments{
Free parameters of fit are shown with
90\% errors ($\Delta \chi^2=2.7$). Fixed parameter values are followed by an 
asterisk (*). $K$: Overall normalization factor for each detector/observation.
Detectors are identified with a subscript (G:GIS S:SIS P:PSPC). 
The normalization is expressed as a 0.5-2 keV flux (unabsorbed) in units of 
${\rm [10^{-12}erg\,s^{-1}\,cm^{-2}]}$. Model Components-- {\bf PL}: 
Power-law with a photon index of $\Gamma$ and a 0.5-2 keV  flux 
of $F_{\rm p12}$. 
{\bf Thermal}: Thermal plasma using 
the XSPEC {\em mekal} model with a plasma temperature $kT$ [keV], 
a metal abundance $Z$ in solar units, \chg{where the solar abundance
is from \citet{angr}}, and a  0.5-2 keV flux of $F_{\rm t12}$. \chg{For Model
J3, the two thermal components are notified by subscripts  
h (hard) and s (soft) respectively. {\bf Disk BB}: The multicolor 
disk \citep{mitsuda} (the XSPEC model {\em diskbb}) with an innerdisk 
temperature $kT_{\rm in}$ and 0.5-2 keV flux of $F_{\rm m12}$.
{\bf Abs:} Absorption by neutral gas 
using the XSPEC model {\em wabs} \citep{wabs} with hydrogen column 
density $N_{\rm H20}$ $[10^{20}{\rm cm}^{-2}]$.}
}
\end{deluxetable}

\end{document}